\documentclass[conference,10pt]{IEEEtran}

\usepackage{amsmath,graphicx,subfigure,epsfig,amssymb,amsmath,epstopdf,tikz,float,tikz-3dplot,color,pdfpages,cite,booktabs,psfrag,multicol}
\usepackage{import}
\usepackage{dsfont}
\usepackage{url}	
\usepackage{algorithmic}
\usepackage{algorithm}
\usepackage{multirow}
\usepackage{graphicx}
\usepackage{geometry}
\usepackage[T1]{fontenc}
\usepackage{comment}
\usepackage[T1]{fontenc}

\usepackage{fancyhdr}

\usepackage{textcomp}
\usetikzlibrary{decorations.pathreplacing,decorations.pathmorphing,decorations.markings,shapes,backgrounds,patterns,decorations.text}

\title{{\fontsize{8pt}{8pt}\selectfont\textcolor{blue}{This is a preprint version. The published version will be available in the proceedings of IEEE FNWF 2025 on IEEEXplore subsequently.}} \\ Digital Twin Assisted Proactive Management in Zero Touch Networks}

\author{Tamizhelakkiya K$^{1}$, Dibakar Das$^{1}$, Komal Sharma$^{2}$, Jyotsna Bapat$^{1}$, and Debabrata Das$^{1}$  \\
$^{1}$Networking and Communication Research Lab,  IIIT Bangalore, India\\
$^{2}$Toshiba Software (India) Private Limited, Bangalore, India}

\providecommand{\keywords}[1]{\textbf{\textit{Keywords---}} #1}

\begin{document}

 \maketitle

\begin{abstract}
The rapid expansion of cellular networks and rising demand for high-quality services require efficient and autonomous network management solutions. Zero Touch Network (ZTN) management has emerged as a key approach to automating network operations, minimizing manual intervention, and improving service reliability. Digital Twin (DT) creates a virtual representation of the physical network in real-time, allowing continuous monitoring, predictive analytics, and intelligent decision-making by simulating what-if scenarios. This paper integrates DT with ZTN proactive bandwidth management in end-to-end (E2E) next-generation networks. The integrated architecture applies Few-Shot Learning (FSL) to a memory-augmented Bidirectional Long Short Term Memory (BiLSTM) model to predict a new network state to augment the known and trained states. Using Q-learning, it determines the optimal action (e.g. traffic shaping) under varying network conditions such that user Quality of Service (QoS) requirements are met. Three scenarios have been considered: 1) normal ZTN operation with closed-loop control, 2) a what-if scenario of DT, and 3) network state unknown to DT. The simulation results show that the network can adapt to underlying changing conditions. In addition, DT-assisted ZTN achieves better performance than the other techniques.

\end{abstract}
\section{Introduction}

The rapid evolution of next-generation communication networks \cite{Kuru2022} has resulted in highly complex and heterogeneous mobile networks, driven by extensive IoT connectivity, diverse services, and advanced architectures. Traditional management and orchestration (MANO) approaches are inadequate to address the dynamic nature and scale of these networks \cite{de2021survey}. To meet the increasing demand for agility, automation, and performance, the Zero-Touch Service Management (ZSM) concept was introduced by the European Telecommunications Standards Institute (ETSI) group \cite{ETSI2019}. It leverages Artificial Intelligence (AI) for real-time decision-making, closed-loop automation to ensure end-to-end (E2E) service orchestration, and improves Quality of Service (QoS) across heterogeneous environments. This framework encompasses various operational functions, such as planning, deployment, delivery, resource provisioning, monitoring, and optimization, all performed autonomously without human intervention. Therefore, it is essential to explore network automation enablers that can complement ZSM to achieve greater automation and operational efficiency. One promising solution is the use of Digital Twins (DTs), to effectively implement and realize the ZSM framework \cite{Khan2022}.

DT technology has emerged as a transformative solution that enables the creation of virtual replicas of physical network environments for real-time monitoring, simulation, and predictive analytics. The integration of DT with AI-driven decision-making, such as the closed-loop Zero Touch Network (ZTN)  \cite{Tamizh2025}. It ensures that networks can self-adapt to changing conditions proactively, minimizing the need for time-consuming manual intervention, and improving overall network efficiency.
The DT framework establishes a bidirectional link between the physical network and a digital domain responsible for network optimization, configuration tuning, AI-driven learning, and predictive analytics \cite{saqib2024digital}. By simulating real-world conditions, DTs enable what-if scenario testing to evaluate new services prior to deployment \cite{kumar2024digital}.

As discussed in \cite{Coro2022}, AI-driven DT allows networks to learn and optimize decision-making processes prior to real-world execution and reduces manual intervention. It closely aligns with and reinforces the principles of ZTN management \cite{Chergui2022}. 
The integration of Deep Learning (DL) and Reinforcement Learning (RL), within DT framework, was shown to dynamically adapt to changing network conditions, ensuring efficient resource utilization and improved system performance \cite{ma2022digital}. In addition, the potential benefits of DT in combination with Open RAN (O-RAN) were explored to improve adaptability, intelligence, and fault tolerance for next-generation networks \cite{Masaracchia2023}. In \cite{Nguyen2025}, the authors illustrated two conceptual frameworks for automating and optimizing O-RAN operations in three different scenarios, which improves operational efficiency. Furthermore, closed-loop mechanisms assisted by DT were employed for energy-efficient slice management in 5G Fixed Wireless Access (FWA) \cite{Ndi2025}.


In \cite{Tamizh2025}, a novel closed-loop control system for ZTN was proposed using XGBoosted Bidirectional Long Short Term Memory (BiLSTM) for traffic prediction. The Q-learning for autonomous decision-making, allowing efficient and proactive adaptive network management. Integrating these ideas with DT has the potential to further enhance network performance. This paper presents an integrated DT-assisted ZTN framework, where a DL model, comprising of Few-Shot Learning (FSL) based memory-augmented BiLSTM, predicts the network state (under bandwidth variation). The Q-learning agent decides actions (e.g. traffic shaping) based on the predicted states to meet E2E QoS requirements. The proposed framework is evaluated through three scenarios explained below.
\begin{enumerate}
\item \textit{Default Case}: The DL model predicts the bandwidth in this baseline scenario, while the Q-learning agent determines the best actions to experience the optimal bandwidth based on the predicted state. This establishes a normal scenario in which the proposed integrated DT assisted ZTN framework operates under predefined conditions.

\item \textit{What-If Scenario}: This scenario introduces a new state that is not known to the network but experimentally found by the DT. The objective of the scenario is to decide on the actions proactively and save them when the network condition actually occurs.

\item \textit{Adaptive Decision-Making Scenario}: This is the scenario when the new state experienced by the physical network but unknown to the DT. It consists of two parts:
\begin{itemize}
\item When a new state is encountered for the first time, the integrated DT assisted ZTN framework initially may choose a suboptimal action (provided by the DT based on past behavior of the network) due to its lack of prior knowledge about the new state .

\item At later instances, the integrated DT assisted ZTN framework learns from simulations of the DT and selects the optimal action when the network experiences the same condition again, showcasing the framework’s ability to adapt and improve over time.

\end{itemize}

\end{enumerate}


Results using \textit{OpenAirInterface (OAI)} \cite{oai} show that the integrated DT-assisted ZTN framework achieves better performance, nearing the ideal network efficiency under bandwidth variation, compared to using only traffic shaping (TS), as well as a combination of TS and ZTN closed-loop control. Thus, the proposed framework connects physical and virtual network environments, facilitating proactive network management, enhancing decision-making, and perpetually optimizing network performance. This study also demonstrates the ability of DT to simulate, predict, and optimize network performance through three specific use cases, advancing the capabilities of the ZTN closed-loop implementation. To the best of our knowledge, none of the previous work has applied the proposed integrated approach of ZTN and DT for network optimization.

The remainder of this paper is organized as follows. Section II introduces the system model. Section III explains the various models applied here. Section IV presents the results and provides a comprehensive discussion. Finally, Section V concludes the paper with future research.

\section{System Model}
\begin{figure}[ht!]
\centering
\includegraphics[width=\columnwidth]{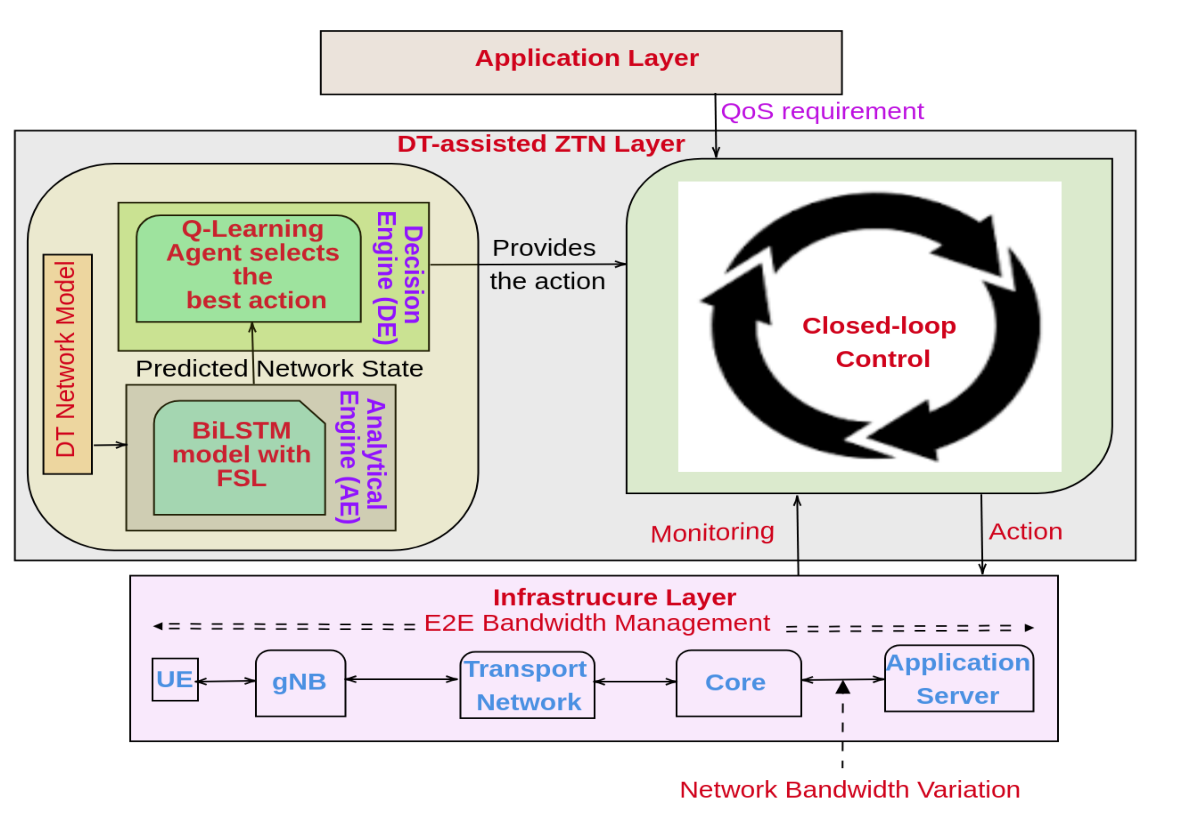} 
\caption{DT assisted ZTN framework for bandwidth variation
control}
\label{congestion_control_DT}
\end{figure}

The system architecture consists of three layers, 1) the application layer, which provides the QoS requirements from the user, 2) the DT assisted ZTN closed-loop control layer, and 3) the physical network infrastructure layer (Fig. \ref{congestion_control_DT}).

\subsection{Application layer}
The application layer provides the QoS requirements of the user, which may be defined by a service level agreement (SLA). To satisfy these requirements, the integrated DT assisted ZTN framework must adapt network policies/actions to handle anomalies like network bandwidth variation.



\subsection{DT assisted ZTN closed loop control layer}
The DT serves as a virtual representation of the network. It incorporates a network state (e.g., bandwidth, data rate, latency, jitter, and packet loss) prediction model that uses a FSL with memory memory-augmented BiLSTM model. The Decision Engine (DE) uses the predicted state as input to a Q-learning agent. The Q-Learning agent facilitates dynamic decision making by evaluating multiple strategies (e.g., throughput control using traffic shaping mechanisms) and selecting the most optimal action based on current network conditions. 

\subsection{Physical network infrastructure layer}
The physical network is a typical Fifth Generation (5G) network with user equipment (UE) served by next-generation node B (gNB), transport network, and the core network (CN). UE is supported by application servers for E2E applications. The physical network provides data about its health, such as bandwidth, signal strength, traffic metrics, power levels, hardware status, etc., to the DT. These metrics ensure that the DT accurately mirrors the network state, enabling predictive maintenance, performance optimization, and consistent synchronization.

\subsection{Memory augmented BiLSTM based prediction system}
The network state prediction is performed using BiLSTM using historical data \cite{Tamizh2025}. The new states are trained using FSL. In this way, the network state prediction model is continually trained \cite{wei2024}. The model is designed to predict the new network state using a \textit{Memory Module} with a \textit{BiLSTM model}, allowing it to dynamically adapt and improve its predictions over time. A dictionary-based memory mechanism has been implemented to store previously encountered sequences and their corresponding outputs \cite{Tran2022}. This enables fast retrieval of known sequences, reducing computation overhead, and improving accuracy. 

\subsection{Q-learning based action selection}
Action selection is performed using Q-learning with the state space, action space, and reward function explained below.
\begin{itemize}
\item \textbf{State Space (\textit{S}):} The E2E bandwidths are the network states predicted by the BiLSTM model.
\item \textbf{Action Space (\textit{A}):} A set of possible E2E bandwidth allocations from the application servers to the UEs applying traffic shaping \cite{Rint2022} parameters to adjust to the underlying link bandwidth (for example, when bandwidth variation is inserted into the User Plane Function (UPF)). The actions involve incrementally increasing or decreasing the allocated E2E bandwidth to closely match the underlying network capacity.
\item \textbf{Reward Function (\textit{R}):} The Q-learning agent receives rewards based on how well the selected action for the E2E bandwidth meets the application requirement in the midst of underlying network bandwidth variation.
\end{itemize}

All optimal actions found by the Q-learning agent in the integrated DT-assisted ZTN framework are stored in the database for future use.

\subsection{Interaction between BiLSTM, Q-learning, and DT models}
\begin{figure}
    \centering
    \includegraphics[width=\columnwidth]{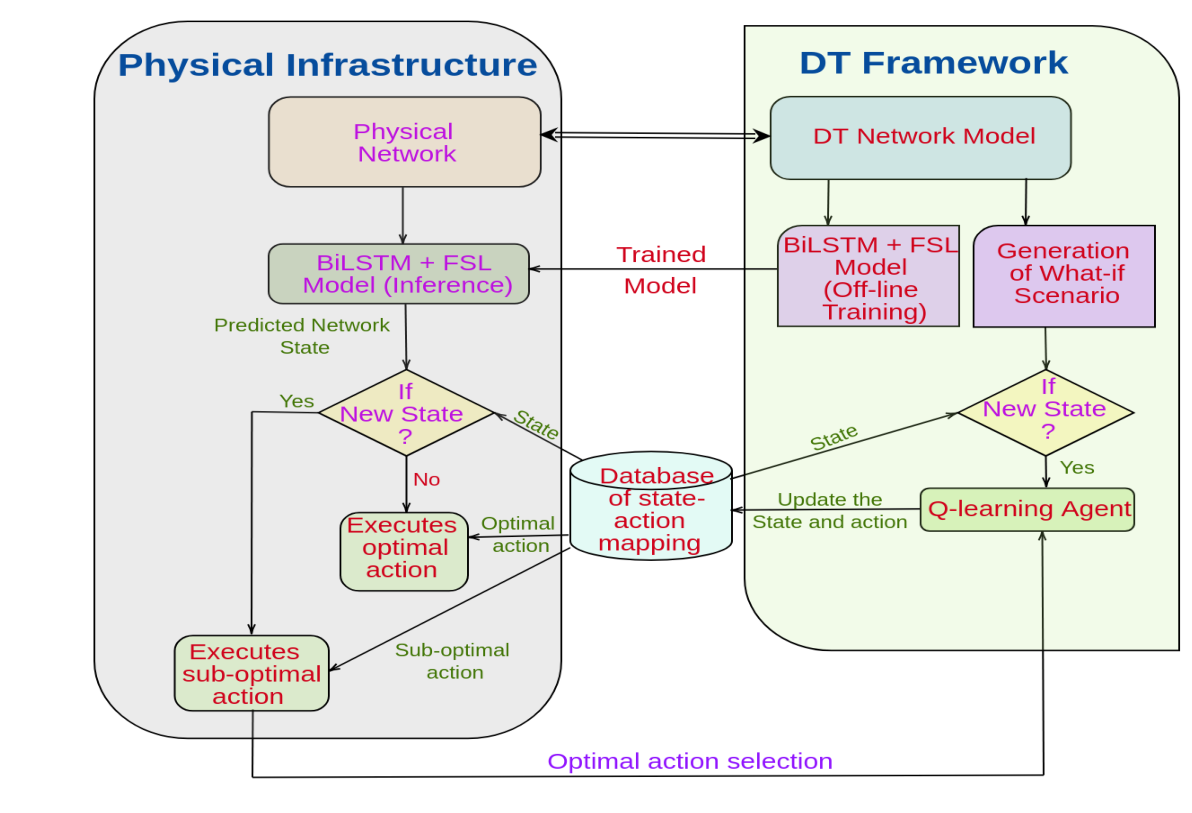}
      \caption{It}
    \label{DT-ZTN-method}
\end{figure}

The interaction between components follows a structured process, as shown in Fig. \ref{DT-ZTN-method}. The FSL based memory augmented BiLSTM model predicts future bandwidth, and the Q-learning agent allocates E2E bandwidth based on these predictions as actions.


\section{Model Details}
\subsection{Traffic generation model}


The traffic generation process assumes that the number of occurrences of data transmissions with a certain rate from the application server to a UE in a time interval follows a Poisson process \cite{Rappaport1996}. The probability of generating $t$ transmissions in a time is given by:
\begin{equation}
P(T = t) = \frac{\lambda^t e^{-\lambda}}{t!}, \quad t = 0, 1, 2, \ldots
\end{equation}

where $T$ is the random variable that represents the number of UE transmissions, $\lambda$ is the average rate (mean number of events per unit time), and  $e$ is Euler’s number ($\approx 2.71828$). Each sampled $t_t \sim \text{Poisson}(\lambda)$ is assigned a bandwidth usage level $b_t$ according to equation \ref{bt}. To train Bi-LSTM, we normalize the raw bandwidth values to a fixed range (usually [0, 1]) using Min-Max normalization (equation \ref{xt}).
     \begin{equation}\label{bt}
     b_t = t_t \times \text{unit\_size}, \quad \text{e.g., unit\_size} = 50, 100, 200, \cdots
     \end{equation} in Kbps.


     \begin{equation}\label{xt}
     x_t = \frac{b_t - b_{\min}}{b_{\max} - b_{\min}},
     \end{equation}

    where, $b_t$ is the bandwidth value at time step $t$. $b_{min}$ is the minimum bandwidth value in the entire dataset (i.e., across all time steps). $b_{max}$ is the maximum bandwidth value in the dataset (Table \ref{maths_parameters}).

The resulting normalized sequence is:
     \begin{equation}
    \mathbf{X} = \{x_1, x_2, \ldots, x_N\}, \quad \text{where each } x_t \in [0,1]
     \end{equation}

\vspace{-0.6cm}

\begin{table}[ht!]
\centering
\scriptsize
\caption{Model Parameters for DT assisted ZTN}
\label{maths_parameters}
\begin{tabular}{|c|c|}
\hline
 \textbf{Symbol} & \textbf{Description} \\ \hline
$T$   & Data transmission rate \\ \hline
$\lambda$ & Average arrival rate of data transmission \\ \hline
$t_t$ & Data transmission sample at $t$-th timestep  \\ \hline
$b_t$ & Bandwidth value at $t$-th timestep \\ \hline
$b_{min}$ & Minimum bandwidth value in the dataset \\ \hline
$b_{max}$ & Maximum bandwidth value in the dataset \\ \hline
$x_t$ & Raw bandwidth value at $t$-th timestep \\ \hline
$L$ & Sequence length for BiLSTM model  \\ \hline
 $\mathcal{M}$ & Memory module \\ \hline

$\mathbf{X}_t$   & Normalized input sequence \\ \hline
$\hat{y}_t$  & Predicted next state at $t$-th timestep \\ \hline
${y}_t$ &  Actual next state at $t$-th timestep  \\ \hline
$y_t^{\text{desired}}$ & Desired next state at $t$-th timestep \\ \hline
$N$ & Number of states for Q-learning \\ \hline
$M$ & Number of actions for Q-learning  \\ \hline
$\mathcal{S}$ & State variable for Q-learning \\ \hline
 $a_j$ & $j$-th optimal action for $i$-th state \\ \hline
 $s_i$ & $i$-th state of Q-learning \\ \hline
 $s_w$/ $s_s$/ $s_o$ & what-if state/ sub-optimal state/ optimal state \\ \hline
  $a_w$/ $a_s$/ $a_o$ & action for what-if state / sub-optimal state/ optimal state\\ \hline
\end{tabular}
\end{table}

\subsection{Memory augmented sequence prediction}
Given a normalized input sequence $\mathbf{X} = \{x_1, x_2, ..., x_N\} \in [0, 1]^N$, to predict the next element $x_{t+L}$, the memory-augmented Bi-LSTM model has been adapted for FSL, where $L$ is the fixed sequence length. For each time step $t \in \{1, 2, ..., N-L\}$,  the input subsequence is defined as
\begin{equation}
\mathbf{X}_t = \{x_t, x_{t+1}, ..., x_{t+L-1}\}
\end{equation}

Let $\mathcal{M}$ be a memory module that matches the normalized input sequences ($\mathbf{X}_t$) with the previously seen outputs:

\begin{equation}
\mathcal{M} : \mathbf{X}_t \mapsto y_t \in [0, 1]
\end{equation}
The prediction of FSL ($\hat{y}_t)$ with memory is defined as:


\begin{equation}
\hat{y}_t =
\begin{cases}
\mathcal{M}(\mathbf{X}_t), & \text{if } \mathbf{X}_t \in \mathcal{M} \\
B(\mathbf{X}_t), & \text{otherwise}
\end{cases}
\end{equation}
Here, $B (\mathbf{X}_t)$ represents the prediction of the BiLSTM model ($B$). $\hat{y}_t$ is from memory ($\mathcal{M}$), if the input sequence $\mathbf{X}_t$ is already stored. Otherwise, it is taken directly from the BiLSTM model output $B(\mathbf{X}_t)$.



\subsection{Memory update rule with threshold ($\delta$):} After a prediction $\hat{y}_t$, the accuracy against the desired value $y_t^{\text{desired}}$ has been verified using the absolute prediction error and threshold $\delta$. 
If the prediction error exceeds a threshold $\delta$, the memory is updated as follows. $$
\text{If } |\hat{y}_t - y_t^{\text{desired}}| > \delta, \quad \text{then } \mathcal{M}[\mathbf{X}_t] \gets y_t^{\text{desired}}, \quad \hat{y}_t \gets y_t^{\text{desired}}
$$
\subsection{Q-learning}
The predicted states ($\hat{y}_t$) of FSL are fed with a state $ \mathcal{S}$ = $s_1, s_2, \ldots, s_N$, where each $ s_i $ represents a network throughput level. The action ($\mathcal{A}$) corresponding to each state, i.e., $\mathcal{A} = \{a_1, a_2, \ldots, a_M\}$ has been decided on the basis of the DT database. The optimal action $a_j$ for a given state $s_i$ is selected based on three distinct scenarios, using predefined mappings and learning-based strategies. If the physical network matches a what-if scenario ($s_i = s_w$) already found by the DT, the decision logic selects an action $a_w$ that yields a throughput value, retrieved from the database generated by the DT. In case of a newly observed state ($s_i = s_s$), not known to the DT, a suboptimal action ($a_j = a_s$) is initially chosen during the first occurrence. For subsequent occurrences of this state, the optimal action ($a_j = a_o$) is selected as found by the DT. For all other states, the optimal action is chosen from the set of available actions $a_j$ already known by the ZTN framework.

\begin{figure}[ht!]
    \centering
    \includegraphics[width=\columnwidth]{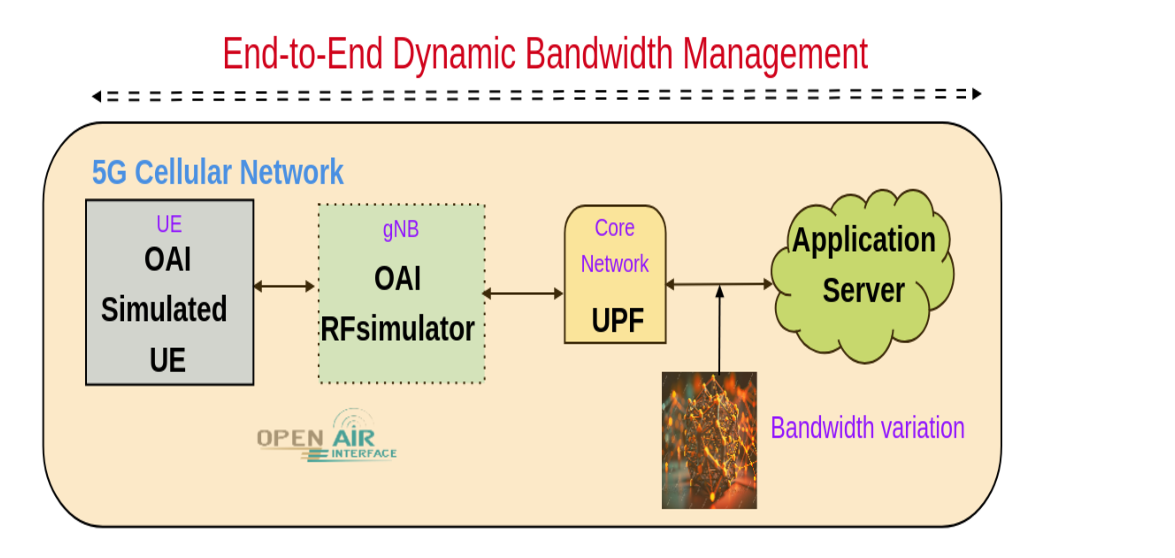}
    \caption{Experimental setup for E2E bandwidth variation management}
    \label{testbed_oai}
\end{figure}
\vspace{-0.4cm}
\section{Results and Discussion}\label{resul_discu}
The 5G cellular system for this work has been developed using the 5G Standalone (SA) framework based on the OAI framework \cite{Goa2021}. This system comprises CN, gNB, and UE as shown in Fig. \ref{testbed_oai}. An external application server simulates an E2E application.  To evaluate performance, E2E downlink (DL) data is generated and the throughput metric is measured using the \texttt{iperf} command. The bandwidth variation is injected between the UPF and the application server, and the E2E bandwidth is adjusted accordingly using the closed-loop control of ZTN assisted by DT.


\begin{table}[ht!]
\centering
\scriptsize
\caption{Simulation parameters}
\label{simulation_parameters}
\begin{tabular}{|c|c|}
\hline
\textbf{Parameter}                                       & \textbf{Value} \\ \hline
Training and testing ambience   & i5 CPU,TensorFlow+Keras \\ \hline
\multicolumn{2}{|c|}{\textbf{OAI 5G-NR platform \cite{oai}}}                       \\ \hline
Number of gNB ($K$) & 1 \\ \hline
Total number of UEs ($N$) & 1 \\ \hline
Channel bandwidth & 40 MHz for DL and UL \\ \hline
Numerology ($\mu$) & 1 \\ \hline
Carrier frequency ($f_c$) & 3.6 GHz \\ \hline
Duplex mode & TDD \\ \hline

\multicolumn{2}{|c|}{\textbf{DT parameters}}                       \\ \hline
Samples in training dataset    & 1800 \\ \hline
Samples in testing dataset  & 700  \\ \hline
Loss function & Mean square error \\ \hline
Optimizer & Adam \\ \hline
Activation function & ReLu \\ \hline
Epochs & 40  \\ \hline
\end{tabular}
\end{table}

\subsection{System configurations and dataset preparation}
\begin{figure}[ht!]
    \centering
     \includegraphics[width=\columnwidth]{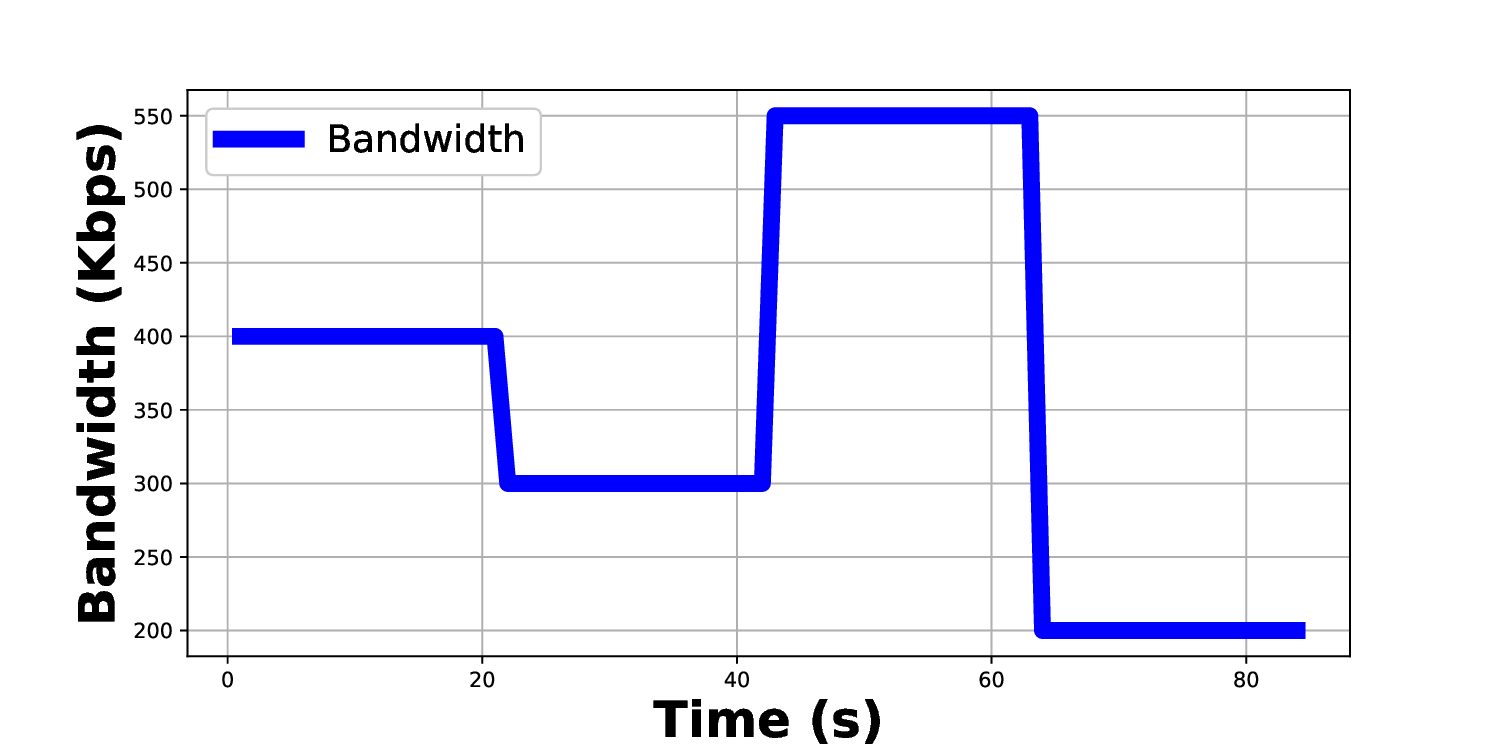}
    \caption{Traffic distribution}
    \label{traffic_pattern}
\end{figure}

The system configuration to develop the 5G SA mode is provided in Table \ref{simulation_parameters}. In addition, it includes the configuration parameters for the DT. The E2E bandwidth (\textit{y}-axis) of a UE follows a Poisson distribution with the mean occurrence rate ($\lambda$) stored as a time series dataset depicted in Fig. \ref{traffic_pattern} over 80 seconds (\textit{x}-axis). Network bandwidth variation scenario is then injected between the application server and UPF. The E2E TS parameters are then configured to mitigate it. These observations are saved and used for model training.


\vspace{-0.5cm}
\subsection{BiLSTM model performance}
\begin{figure}[ht!]
    \centering
         \includegraphics[width=\columnwidth]{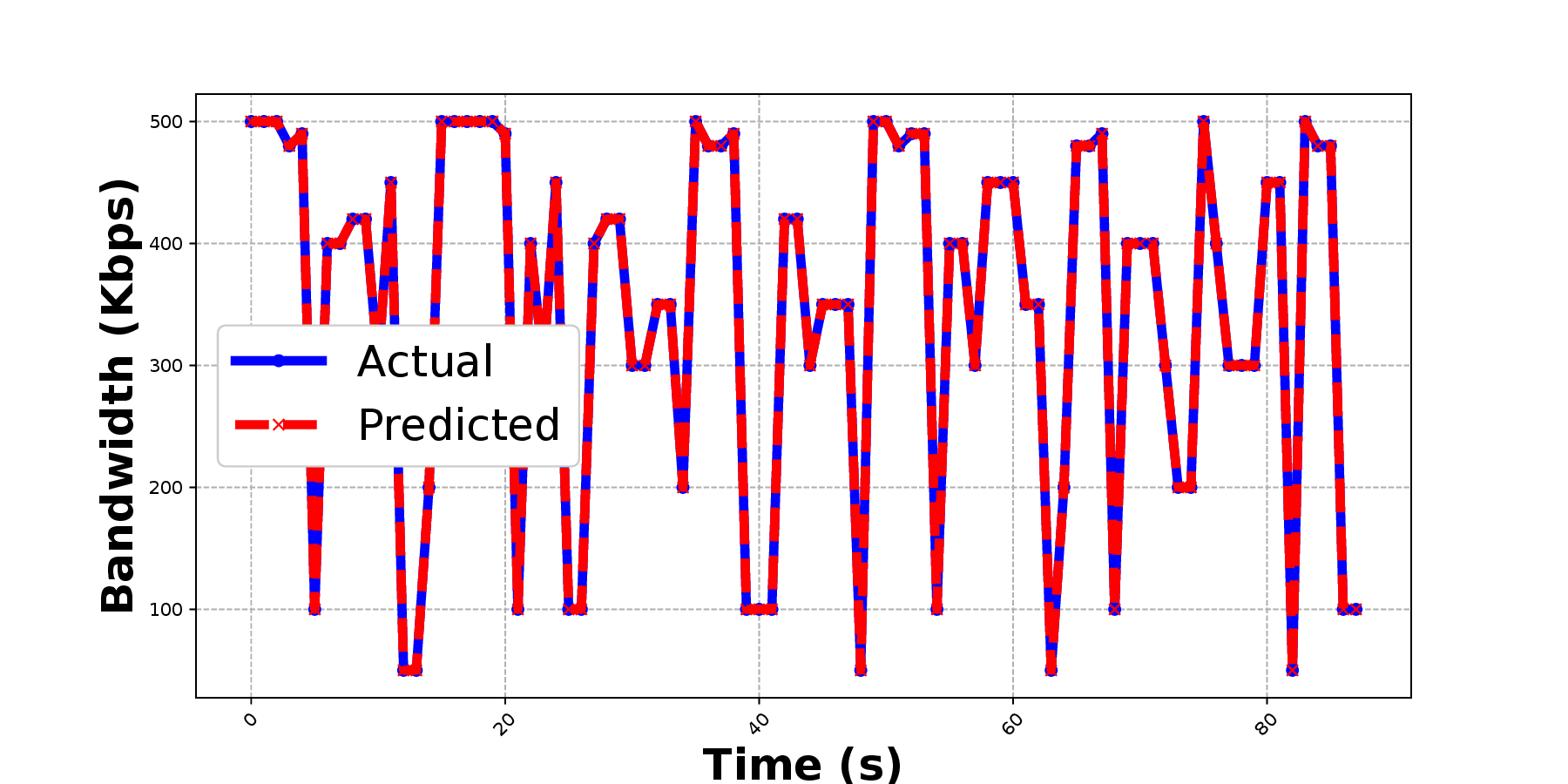}

    \caption{BiLSTM model with FSL traffic prediction performance}
    \label{hybird_per}
\end{figure}

The memory-augmented BiLSTM model is trained through the FSL, and it predicts the next state bandwidth. The predicted E2E bandwidth (red line) has been matched with the actual value (blue line) as shown in Fig. \ref{hybird_per}.

\begin{figure}[ht!]
\centering
\subfigure[]{\includegraphics[width=\columnwidth]{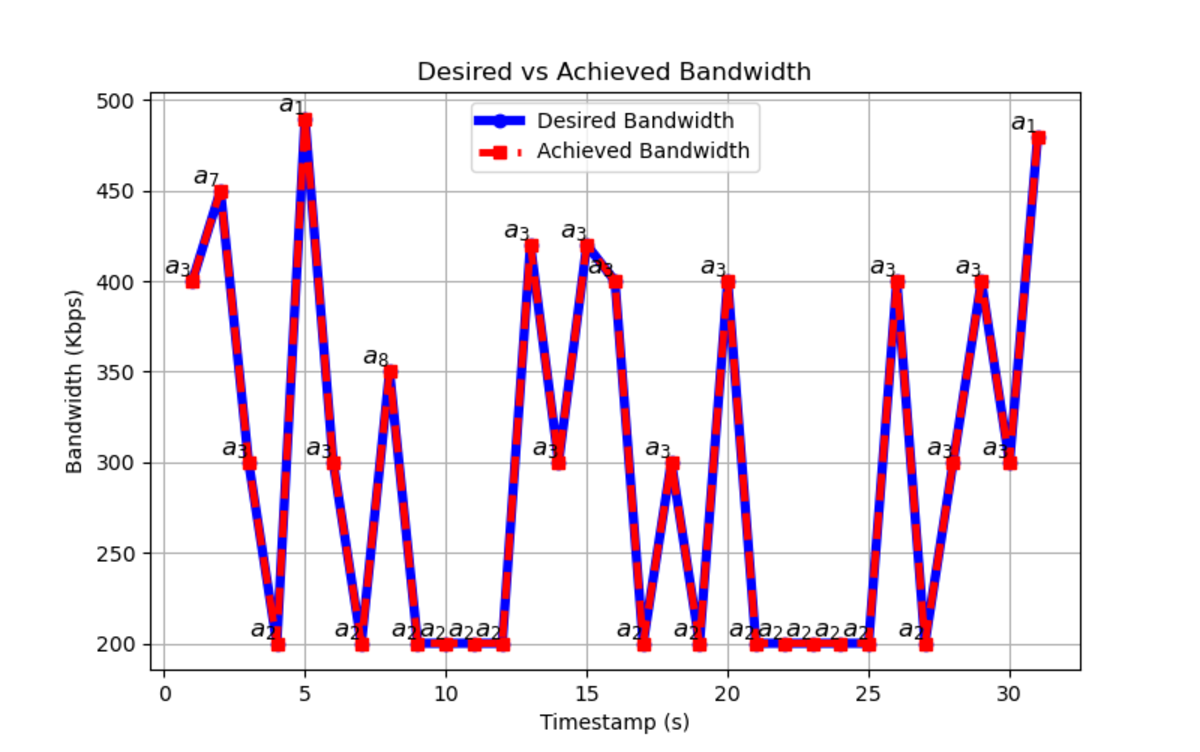} \label{traffic_default}}
\hfil
\subfigure[]{\includegraphics[width=\columnwidth]{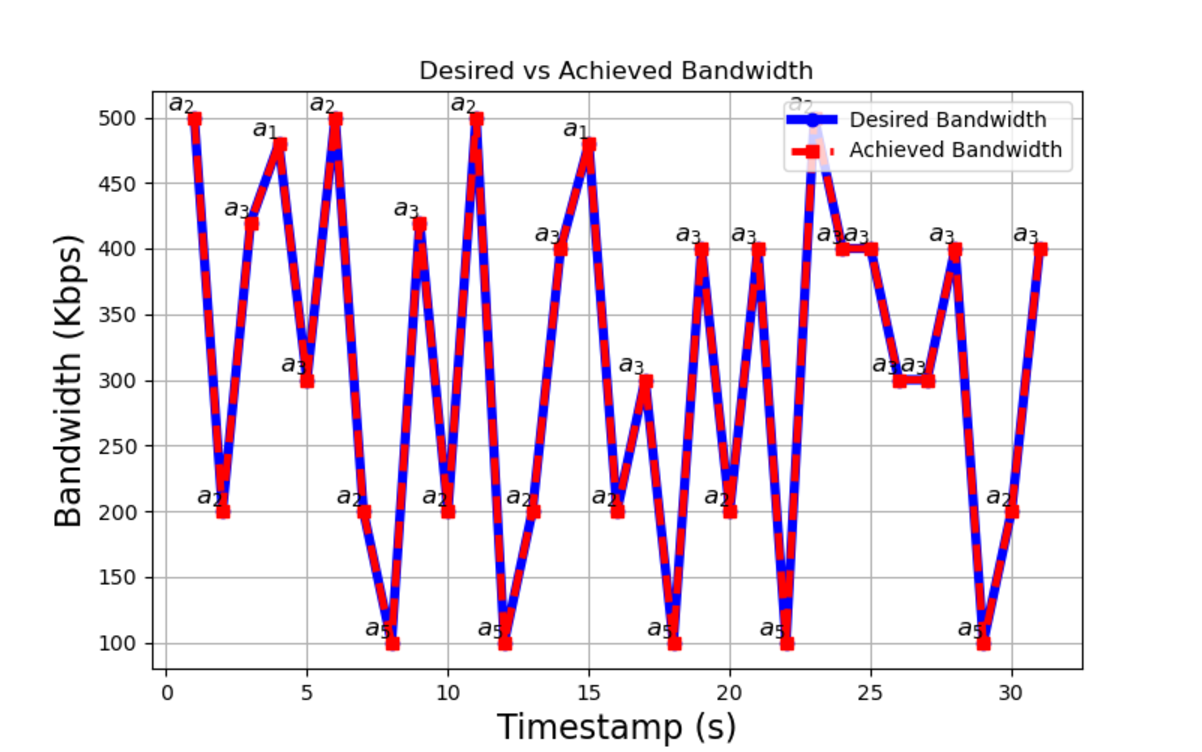}\label{traffic_whatif}}
\caption{Effectiveness of Q-learning agent (a) Default case and (b) What-if case}
\label{DT_1}
\end{figure}
\vspace{-0.5cm}
\subsection{DT performance analysis}
The actions ($a_i, i=1,2,..$) are assigned to the predicted state by the DT model, as illustrated in Fig. \ref{DT_1}.  In the default case, the Q-learning agent has already learned to allocate E2E bandwidth across eight known (network BW variation) states. It has been observed that the achieved bandwidth (red line) closely follows the actual bandwidth trend (blue line) in the default scenario (Fig. \ref{traffic_default}). Furthermore, when a new state of 100 Kbps (along \textit{y}-axis) is introduced as a what-if scenario, the DT uses the Q-learning model to determine the optimal action, ensuring that the desired E2E bandwidth is maintained in this dynamic bandwidth state, as shown in Fig. \ref{traffic_whatif}.

\begin{figure}[ht!]
\centering
\subfigure[]{\includegraphics[width=\columnwidth]{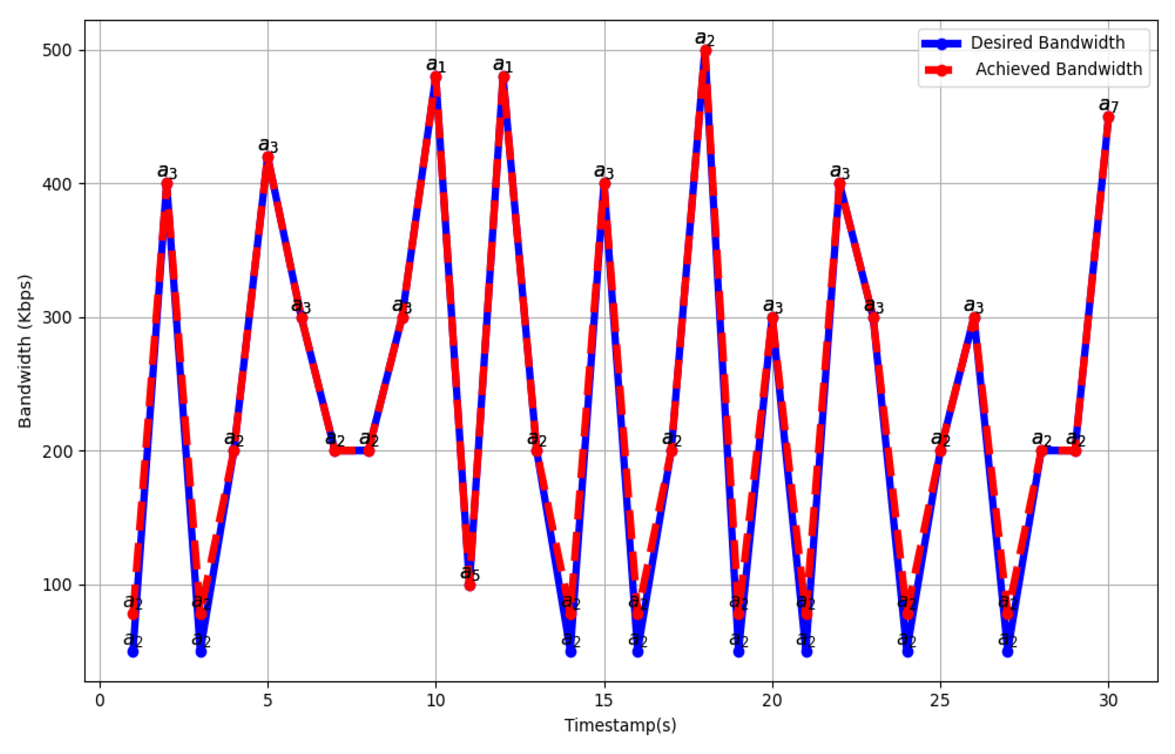}\label{traffic_subopti}}
\hfil
\subfigure[]{\includegraphics[width=\columnwidth]{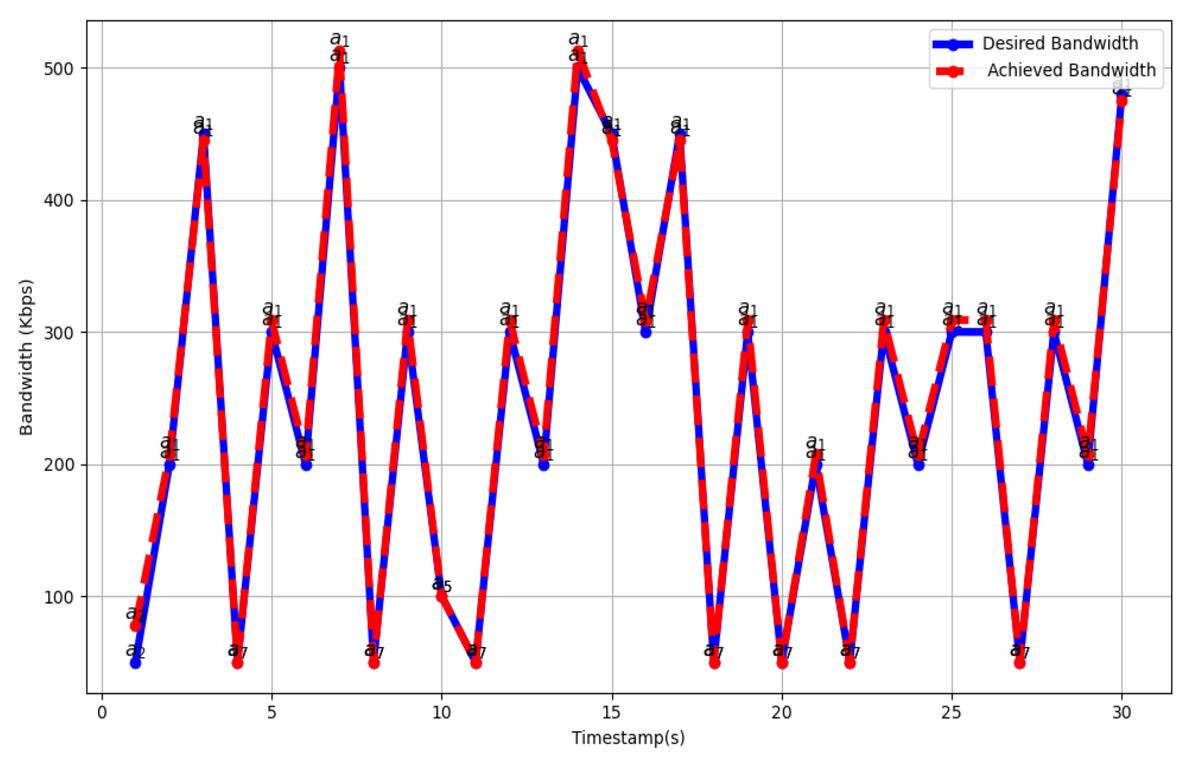}\label{traffic_opti}}
\caption{{Effectiveness of Q-learning agent (a) Suboptimal case and (b) Optimal case}}
\end{figure}

Additionally, the adaptive scenario (physical network observes a state which is not known to DT) examines how the DT-assisted ZTN framework initially selects a suboptimal action for a new state, but gradually learns to choose the optimal action in subsequent occurrences. We analyze adaptive decision-making by introducing another new state of 50 Kbps. Initially, the DT model selects a suboptimal action of 70 Kbps (Fig. \ref{traffic_subopti}) when this state first appears. However, at a later point in time, the DT model is adjusted and provides an optimal action of 50 Kbps, as depicted in Fig. \ref{traffic_opti}. Thus, this framework ensures proactive and intelligent bandwidth allocation, bridging the gap between real and virtual environments while improving resource management in 5G networks.

\vspace{-0.5cm}
\subsection{Throughput evaluation under bandwidth variation using TS, ZTN, and DT techniques}
\begin{figure}[ht!]
\centering
\includegraphics[width=\columnwidth]{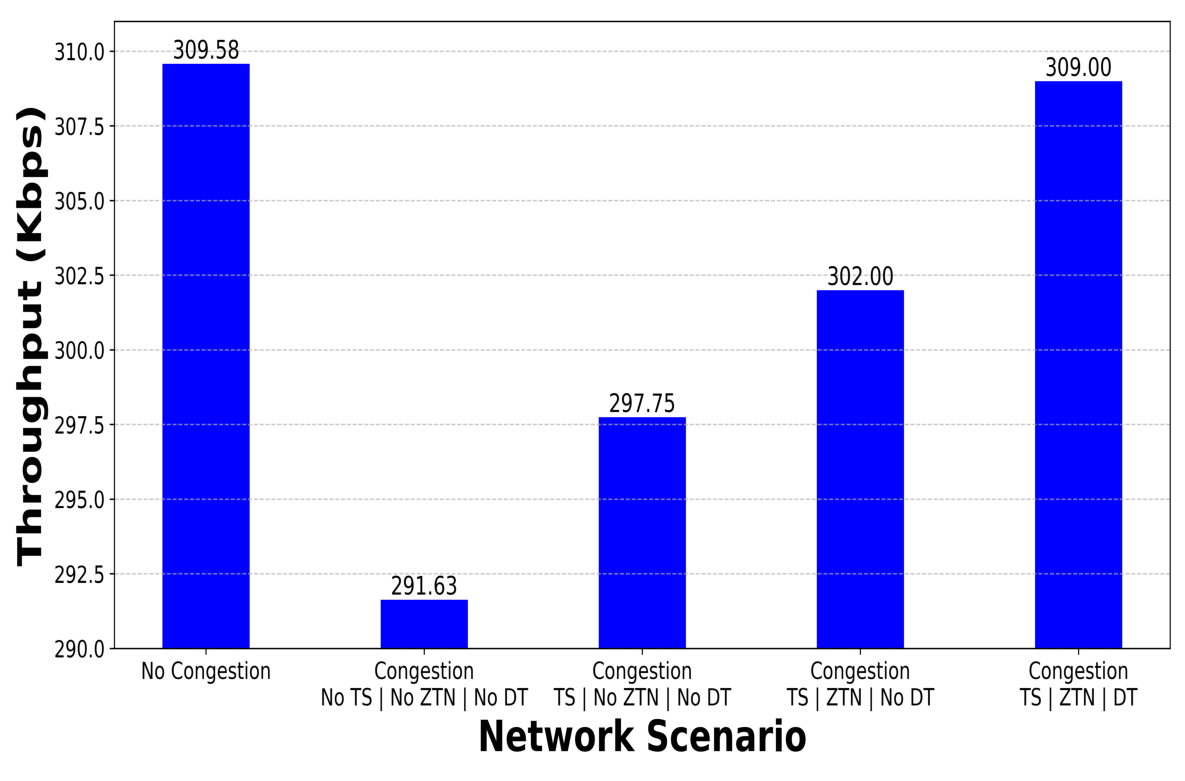}
\caption{Performance comparison of different techniques }
\label{bandwidth_differentcases}
\end{figure}

The average E2E throughput between UE and the application server was measured over a period of time to meet a user required bandwidth of $\approx$ 310 Kbps in five different network scenarios: 1) no bandwidth variation, 2) with bandwidth variation, 3) with bandwidth variation combined with TS, 4) with bandwidth variation + TS + ZTN and 5) with bandwidth variation + TS + ZTN + DT (Fig. \ref{bandwidth_differentcases}).

It has been observed that throughput decreased with increased network bandwidth variation. To mitigate this, TS techniques were introduced in traffic between UE and the application server, which improved throughput, but did not meet the desired user QoS requirement. The ZTN closed-loop mechanism was then implemented alongside TS, increasing throughput to approximately 302 Kbps. Combining the DT approach with TS and ZTN allowed the system to reach throughput levels close to an ideal (no bandwidth variation) scenario, achieving approximately 309 Kbps. This improvement demonstrates the effectiveness of integrating strategies using TS, ZTN, and DT together to optimize network performance under varying network conditions.

\section{Conclusion}
Integrating DT technology with ZTN offers a novel approach towards proactive management in cellular networks. By utilizing real-time monitoring, AI-driven predictive analytics, and closed-loop optimization, this framework improves network adaptability, resource allocation, and ability to meet user QoS requirements. The DT’s capability to create a virtual representation of the physical network and simulating different scenarios facilitates intelligent and proactive decision-making. Adopting FSL integrating with memory-augmented BiLSTM ensures robust and adaptive prediction of network state. This helps in improved ZTN closed-loop functionality towards applying optimal actions such that user QoS requirements are met. The results show the overall effectiveness of this integrated framework in reducing the variation of network bandwidth and meeting the user QoS requirements from an E2E perspective.

Future work will try to extend this integrated framework to a general setting with different network parameters beyond bandwidth variation control techniques.



\section*{Acknowledgment}
The authors would like to thank Toshiba Software India Pvt. Ltd. for sponsoring this research project.

\bibliographystyle{IEEEtran}
\bibliography{References.bib}

\end{document}